\documentclass[12pt]{article}
\usepackage{graphicx}
\usepackage{subfigure}
\title{\textbf{Experimental treatment of Quark and Gluon Jets}}

\author{So\v na Pochybov\'a $^{1,2}$ \\
	\small{$^{1}$ E\"otv\"os L\'orand University, Pazm\'any P\'eter s\'et\'any 1/A, H-1117, Budapest, Hungary}\\ 
	\small{$^{2}$ MTA KFKI RMKI, Konkoly-Th\'ege Mikl\'os \'ut 29-33, H-1121, Budapest, Hungary}\\ 
	\small{\itshape sona.pochybova@cern.ch}
	}
\begin{document}


\maketitle
\begin{abstract}
The separate study of quark and gluon jets is vital for the interpretation of multiple variables behaviour observed in both high-energy hadron and heavy-ion collisions in the present and future experiments.
We propose a set of jet-energy dependent cuts to be used to distinguish between quark and gluon jets experimentally based on a Monte-Carlo study of their properties. Further, we introduce the possibility to calibrate these cuts via gamma-jet and multi-jet events, which represent clean production channels for quark and gluon jets, respectively. The calibration can happen on real data and thus, reduces the dependence of the method performance on Monte-Carlo model predictions.
\end{abstract}

\section{Introduction}
\label{Sec:Sec1}
Jets are objects produced in hard scatterings  of  colliding particles. Experimentally we can observe jets as showers of high momentum particles. The character of these showers is determined by the fragmentation properties of the original parton; quark or gluon. In QCD, quarks and gluons carry different color factors \cite{Ellis:1991qj}. This factor is proportional to the probability of a parton to radiate a soft gluon. Gluons have more than twice the color factor as quarks and as such are expected to form broader and higher multiplicty jets with softer fragmentation function.

Apart from these differences, the gluons are expected to contribute significantly to the baryon production as compared to quarks \cite{Albino:2005me}. All these differences naturally must be demonstrated in the particle spectra observed in an experiment. Previous experiments with $e^+e^-$ \cite{Abreu:1995hp} and $p\bar{p}$ \cite{Affolder:2001jx} collisions studying the properties of different parton types have qualitatively proven these expectations.

From  the heavy-ion perspective, the study of fragmentation properties of quarks and gluons becomes important for understanding unexpected observation from RHIC explained by different phenomenological models (e.g. coalescence \cite{Greco:2003xt}, jet flavor conversion \cite{Liu:2007zz}), which incorporate the above mentioned differences. 

Our aim is to perform a systematic study of the baryon and meson production inside quark and gluon jets. For this, we need to identify the jets first. In the following we introduce a data driven method to distinguish quark and gluon jets and make the study of their properties experimentally feasible.

\section{Method description}
\label{Sec:Sec2}

\begin{figure}[ht]
\begin{minipage}[ht]{0.3\linewidth}
\subfigure[] 
{
\includegraphics[scale=0.3]{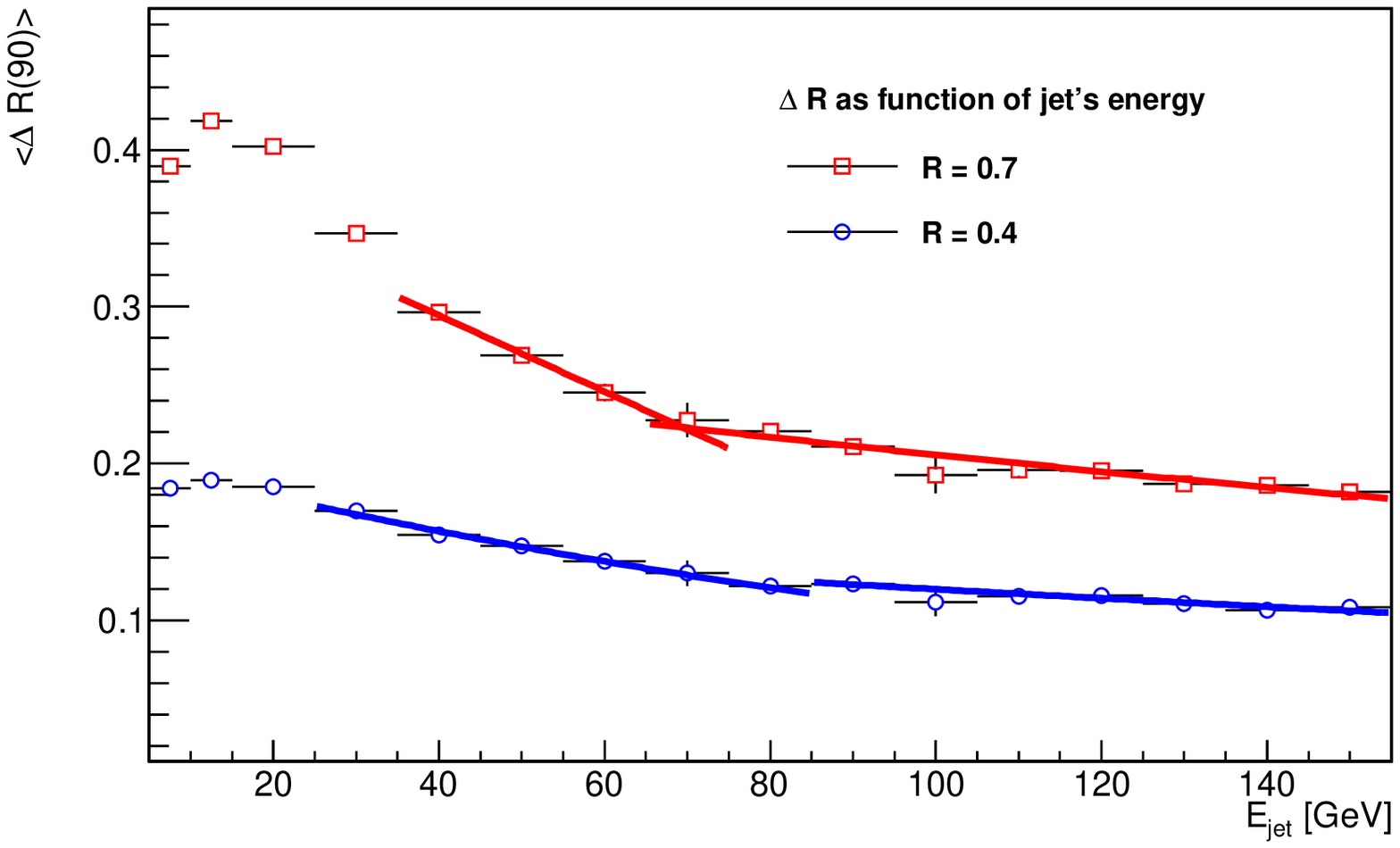}
}
\end{minipage}
\hspace{1cm}
\begin{minipage}[ht]{0.6\linewidth}
\subtable[] 
{
\begin{tabular}{c|c c c c}
\hline
R&$E_{jet}$ int.&Fit fction&A&B\ ($\times\ 10^{-3}$)\\
\hline
$0.4$&$(25;85)$&$exp$&$-1.590$&$-6.556$\\
$0.4$&$(85;155)$&$exp$&$-1.878$&$-2.438$\\
\hline
$0.7$&$(35;75)$&$pol1$&$-0.391$&$-2.424$\\
$0.7$&$(75;155)$&$exp$&$-1.316$&$-2.666$
\vspace{0.2in}
\end{tabular}
}
\end{minipage}
\caption{Panel (a): $<\Delta R(90\%)>$ as function of jet's energy for $R=\{0.4,0.7\}$. Lines represent the fits in separate energy intervals. Panel (b): Table with fit parameters. Indicated are the fit functions for different energy intervals, where $exp$ corresponds to $\Delta R(90\%)_{calc}=exp\{A+B\times E_{jet}\}$ and $pol1$ to $\Delta R(90\%)_{calc}=A+B\times E_{jet}$.}
\label{Fig:1}
\end{figure}
The experimental treatment of quarks and gluons in proton-proton collisions seems challenging since we only can observe the final state hadrons together with the underlying event. Therefore, our observation is restricted to the experimental definition of a jet in terms of jet finding algorithms. Further, the study of jet properties based on their identification may be biased by our prior expectations incorporated into Monte-Carlo models.

However, experimental data offer the possibility to distinguish between quark and gluon in an unbiased way, by observing their properties in channels, were we are certain of the origin of the jet. Such channels are the multi-jet and gamma-jet events, sources to gluon and quark jets respectively. Observing the properties of jets in such events can help us to identify the leading jets in others.

The study was performed on simulations done using the Pythia 6.4 Monte-Carlo generator with the settings of $Perugia0$ tune \cite{Skands:2010ak}. For testing 4 data samples were generated, each containing 1 milion events. The samples involved 3 sets of events with hard scatterings, divided based on the two leading jets into gluon-gluon (GG), quark-quark (QQ) and quark-gluon (QG) samples. In order to study quark jets, gamma-jet sample was created ($\gamma$Q). To reconstruct the jets, we used $\mathrm{anti}$-$\mathrm{k_T}$ jet-finding algorithm \cite{Cacciari:2008gp} for 2 jet-size parameters; $R=\{0.4, 0.7\}$. In order to design the cut, we chose to compare the subcone size, which contains $90\%$ of jet's energy - $\Delta R(90\%)$, after the tracks have been sorted in distance from the jet axis. The method introduced is performed in two steps.

\paragraph{$1^{st}\ step$} 
From each event we select two leading jets and measure their $\Delta R(90\%)$. We plot it as a function o jet's energy (see Fig.\ \ref{Fig:1}, panel (a)). We fit $<\Delta R(90\%)>$ and so we obtain $\Delta R(90\%)_{calc}$ (see Fig.\ \ref{Fig:1}, panel (b)).

\begin{figure}[ht]
\includegraphics[width=1\textwidth]{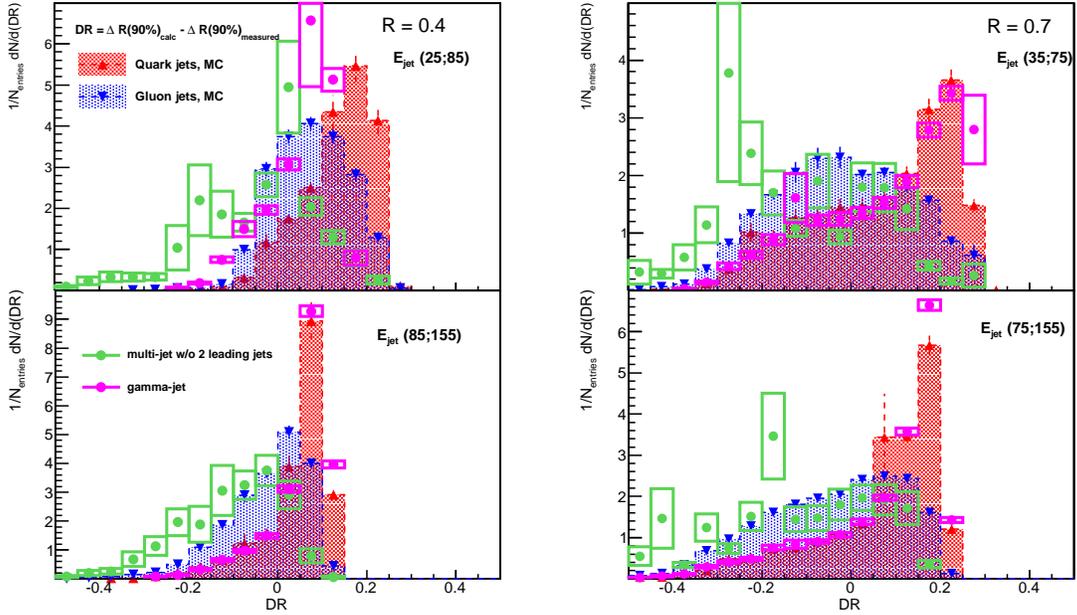}
\caption{$DR$ variable distribution for $R=\{0.4,0.7\}$ and different jet energy intervals for MC quark and gluon jets compared to $DR$ distribution of quark and gluon jets obtained from gamma-jet and multi-jet events respectively.}
\label{Fig:2}
\end{figure}

\paragraph{$2^{nd}\ step$} 
Next we reconstruct jets in multi-jet and gamma-jet events in order to obtain samples of gluon and quark jets. From multi-jet events we select all but the two leading jets, from gamma-jet events we take the jet at $180^o\pm 30^o$ degrees w.r.t. the gamma. We measure $\Delta R(90\%)_{measured}$ for the selected quarks and gluons and check how it is distributed around $\Delta R(90\%)_{calc}$. We obtain a distribution of $DR=\Delta R(90\%)_{calc} - \Delta R(90\%)_{measured}$ plotted in Fig.\ \ref{Fig:2}. Based on this distribution we choose a $DR$ cut to be applied to the leading jets (Table \ref{Tab:1}).

\begin{table}[ht]
\begin{center}
\subtable[]{
\begin{tabular}{c||c c c}
$R=0.4$&$E_{jet}$ int.&$Q_{cut}$&$G_{cut}$\\
\hline
\ &$(25;85)$&$0.15$&$0.10$\\
\ &$(85;155)$&$0.05$&$0.00$
\end{tabular}
}
\subtable[]{
\begin{tabular}{c||c c c}
$R=0.7$&$E_{jet}$ int.&$Q_{cut}$&$G_{cut}$\\
\hline
\ &$(35;75)$&$0.15$&$0.00$\\
\ &$(75;155)$&$0.10$&$0.05$
\end{tabular}
}
\caption{Cuts to select quark and gluon jets based on gamma-jet and multi-jet distributions of $DR$ in Fig.\ \ref{Fig:2}.}
\label{Tab:1}
\end{center}
\end{table}

\section{Discussion}
\label{sec:3}
\begin{figure}[ht]
\subfigure[]{
\includegraphics[width=0.5\textwidth]{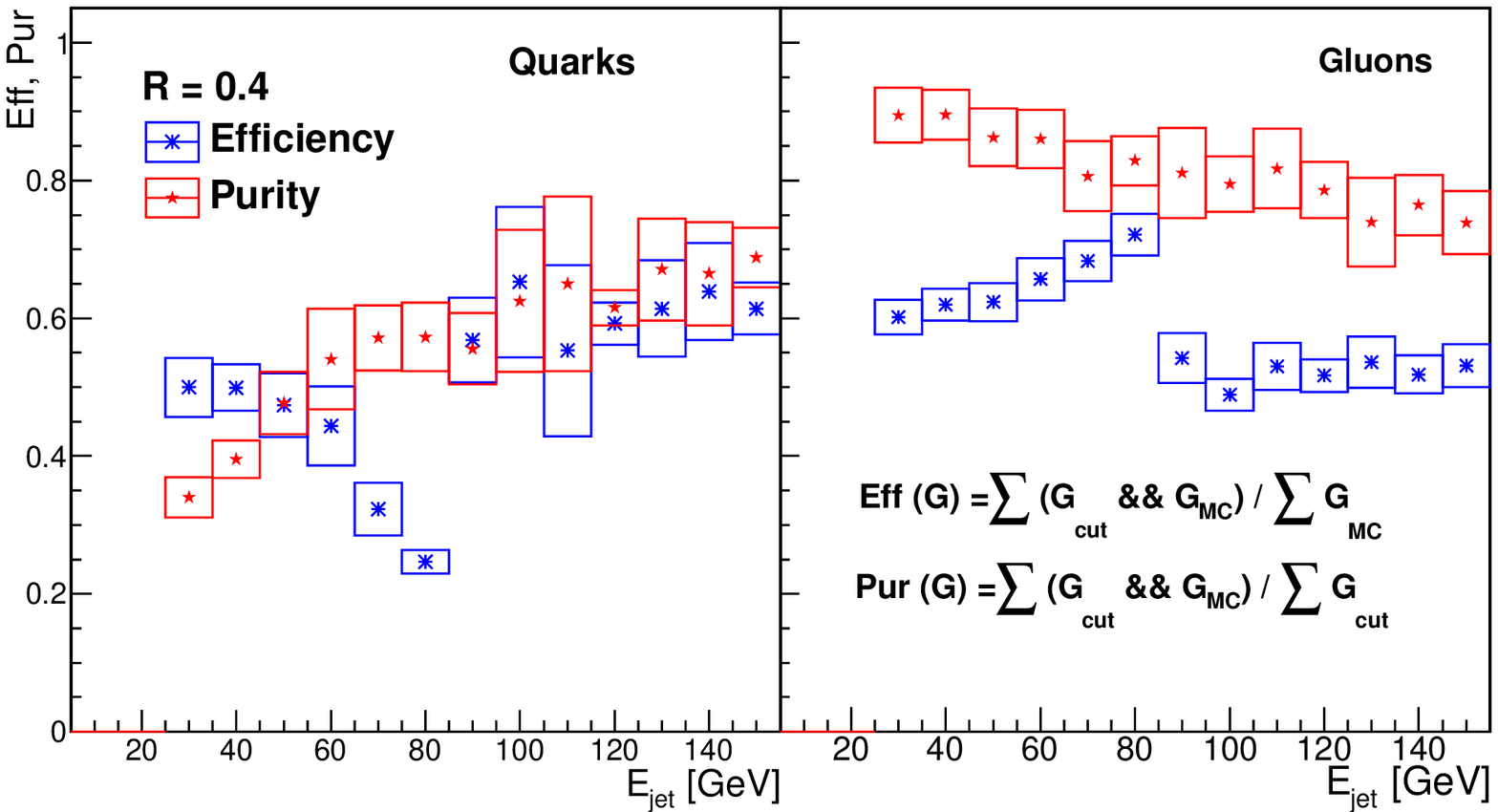}
}
\subfigure[]{
\includegraphics[width=0.5\textwidth]{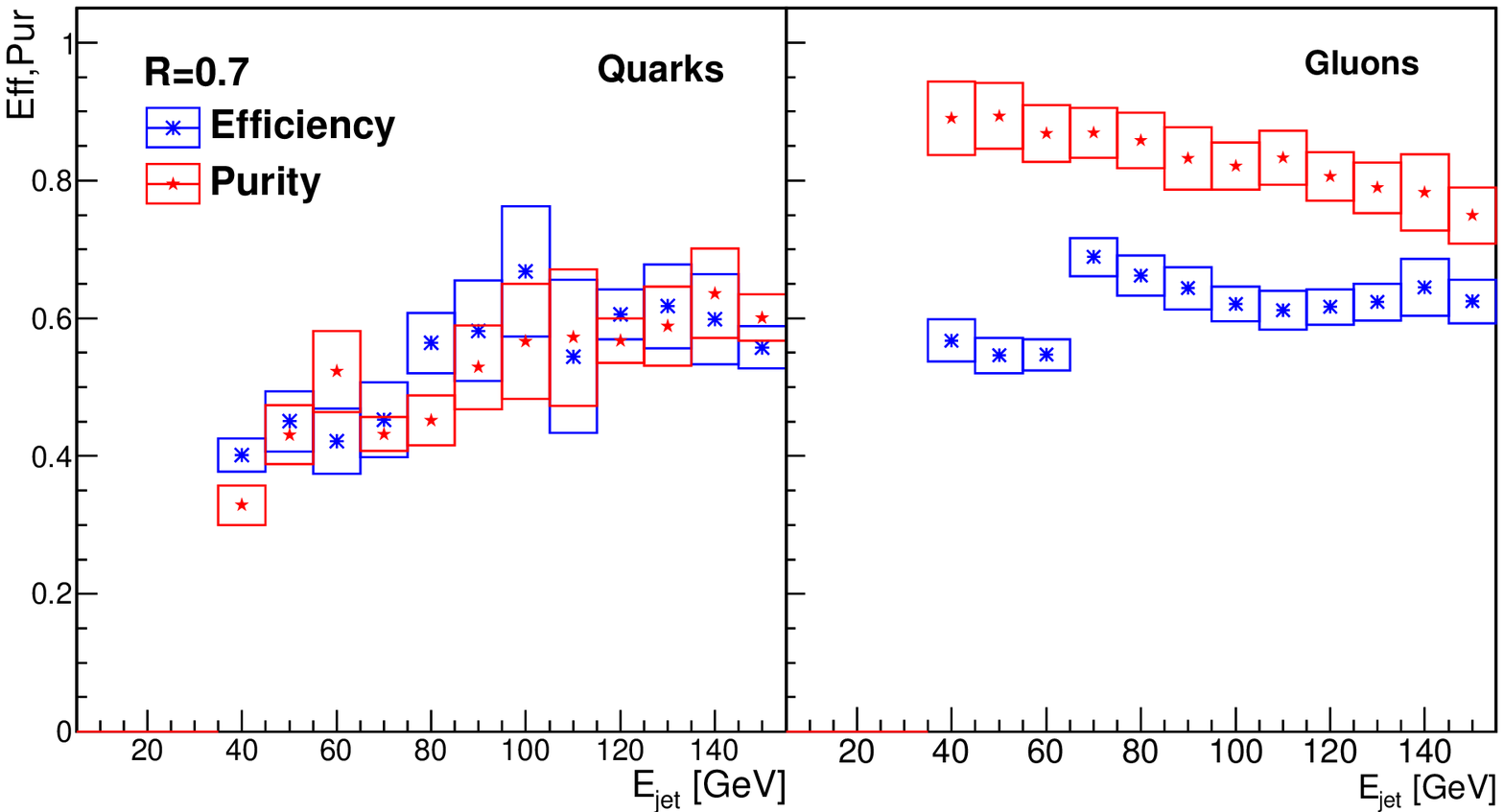}
}
\caption{Performance plots. Efficiency and purity as function of $E_{jet}$.}
\label{Fig:3}
\end{figure}

We presented a possibility to distinguish between quark and gluon jets experimentally. The main advantage of this method lies in the fact, that the cut on the variable introduced can be calibrated on real data without the necessity to rely on Monte-Carlo information. This way, our selection is not biased by our prior expectations about the differences of quark and gluon jets.

The performance of the method was determined by the efficiency and purity of the identification of the two leading jets in an event. As can be seen in Fig.\ \ref{Fig:3}, for quarks it achieves efficiency and purity up to 60\% in the higher energy interval. For gluons, efficiency of the selection is better (constant, 60\% for higher energy interval), although the purity drops slightly with energy. The rise of purity for quarks and the drop for gluons is however expected since quarks are to form harder jets than gluons. The method performs better for bigger jet size, namely $R=0.7$ (see Fig.\ \ref{Fig:3}, panel(b)).

The cuts were applied for two energy intervals. In Fig.\ \ \ref{Fig:2} we see, that the gamma-jet and multi-jet $DR$ distribution does not overlap well with the Monte-Carlo quarks and gluons in the lower energy interval, especially for $R=0.4$, which distorts our selection and worsens the performance. We need to be aware of this and apply a more detailed treatment of this region.

\section*{Acknowledgements}

I would like to thank my supervisor P\'eter L\'evai for cooperation and the support of my ideas. The presentation of this work was possible thanks to the funding provided by ELTE and the Hungarian OTKA 77816.


\end{document}